\documentclass[aps,preprint,epsfig,epsf,superscriptaddress]{revtex4}
\usepackage{graphicx}
\usepackage{amssymb}
\usepackage{amsmath}
\usepackage{times}
\begin{document}

\title{Mass-Transport Models with Fragmentation and Aggregation}


\author {Gaurav P. Shrivastav}
\email{gps.jnu@gmail.com}
\affiliation{School of Physical Sciences, Jawaharlal Nehru University,
New Delhi -- 110067, India}

\author {Varsha Banerjee}
\email{varsha@physics.iitd.ac.in}
\affiliation{Department of Physics, Indian Institute of Technology,
Hauz Khas, New Delhi -- 110016, India.}

\author {Sanjay Puri}
\email{puri@mail.jnu.ac.in}
\affiliation{School of Physical Sciences, Jawaharlal Nehru University,
New Delhi -- 110067, India}


\begin{abstract}
We present a review of nonequilibrium phase transitions in mass-transport models with kinetic processes like fragmentation, diffusion, aggregation, etc. These models have been used extensively to study a wide range of physical problems. We provide a detailed discussion of the analytical and numerical techniques used to study mass-transport phenomena.
\end{abstract}

\maketitle

\section{Introduction}
\label{sec1}

There has been intense research interest in phase transitions in mass-transport and growth models involving adsorption and desorption, fragmentation, diffusion and aggregation. These processes are ubiquitous in nature and arise in a large number of seemingly diverse systems such as growing interfaces \cite{lagally,z&l}, colloidal suspensions \cite{white}, polymer gels \cite{ziff}, river 
networks \cite{sch}, granular materials \cite{beads}, traffic flows \cite{traffic}, etc. 
In these systems, different nonequilibrium states arise if the rates of the underlying 
microscopic processes are varied. Conservation laws also play an important 
role in determining both the time-dependent behavior and steady states of such systems. As these steady states are usually not described by the Gibbs distribution, they are hard to determine.  However, much insight on this issue has been gained by studying lattice models. Due to their simplistic nature, these models can be treated either exactly or via a mean-field (MF) approach \cite{marro,haye,krapiv,sk3,rajesh,lev2004}. They are also simple to implement numerically using Monte Carlo (MC) techniques \cite{bh88,nb99}.

In this paper, we present a pedagogical discussion of the modeling and simulation of mass-transport and growth phenomena. We discuss analytical and numerical techniques in the context of mass-transport models where the elementary move is the fragmentation of mass $k$, and its subsequent diffusion to a neighboring site where it aggregates. The $k$-chip models that we study here are interesting in physical situations where the deposited material consists of polymers. We study the MF limit of these models, focusing on the steady-state mass distribution [$P(m)$ vs. $m$], which is characterized by $k$ branches. We also compare the MF results with MC simulations in $d$ $=$ $1,2$.

This paper is organized as follows. In Sec.~\ref{s2}, we present  a framework for mass-transport models in terms of the rate (evolution) equations for $P(m,t)$, the probability that a site has mass $m$ at time $t$. We then discuss various systems which can be described within this framework. In Sec.~\ref{s3}, we introduce $k$-chip models and obtain analytical results for the MF versions of these models. In Sec.~\ref{s4}, we present MC results for mass-transport models, and compare them with the corresponding MF solutions. We conclude this paper with a summary and discussion in Sec.~\ref{s5}. The appendices contain details of calculations and MC procedures.

\section{Framework and Applications of Mass-Transport Models}
\label{s2}

\subsection{Lattice Models and Rate Equations}
\label{s2a}

We consider lattice models of mass transport with the processes of fragmentation,
diffusion and aggregation. For simplicity, we describe the models on a 1-dimensional 
lattice with periodic boundary conditions. (The generalization to higher dimensions is 
straightforward.) To begin with, masses are placed randomly at each site 
with an overall mass density $\rho$. Let $m_{i}(t)$ be the mass at site $i$ at time $t$. The mass variables assume discrete values 0,1,2,3, etc. 
The evolution of the system is as follows. A piece of mass $n$ chips off a site having mass 
$m~(\geq n)$ with rate $g_{m}(n)$. This piece deposits on the right neighbor with probability $p$,
or on the left neighbor with probability $1-p$. The mass of the chosen neighbor adds up, while 
that of the departure site decreases, with the total mass of the system remaining conserved. 
Figure~1 is a schematic depiction of the above model. To facilitate MC simulations, the update rules can be rewritten as follows: \\
1) Randomly pick a site $i$ at time $t$ with mass $m_{i}(t)=m$. The site is updated as $m_{i}(t+1)= m_{i}(t)-n$ with rate $g_{m}(n)$. \\
2) The neighboring sites are updated as $m_{i+1}(t+1)=m_{i+1}(t)+n$ with probability $p$, or
$m_{i-1}(t+1)=m_{i-1}(t)+n$ with probability $1-p$.

We also study the above models within a MF approximation which keeps track of the
distribution of masses, ignoring correlations in the occupancy of adjacent sites. 
Although the MF theory has this shortcoming, our MC simulations show
that it gives an accurate description of the above model, even in the 1-dimensional case.
Let $P(m,t)$ denote the probability that a site has mass 
$m$ at time $t$. In the MF limit, $P(m,t)$ evolves as follows:
\begin{eqnarray}
\label{rate1}
\frac{d}{dt}P(m,t)&=&-P(m,t)\sum_{m_{1}=1}^m g_{m}(m_{1})
      -P(m,t)\sum_{m_{2}=1}^{\infty}P(m_{2},t)\sum_{m_{1}=1}^{m_{2}}
	g_{m_{2}}(m_{1})\nonumber \\
  & &+\sum_{m_{1}=1}^{\infty}P(m+m_{1},t)g_{m+m_{1}}(m_{1})\nonumber\\
      & & +\sum_{m_{1}=1}^{m}P(m-m_{1},t)\sum_{m_{2}=m_{1}}^{\infty} 
    P(m_{2},t)g_{m_{2}}(m_{1}), \ \ \ m \ge 1,
\end{eqnarray}
\begin{eqnarray}
\label{rate10}
\frac{d}{dt}P(0,t)&=&-P(0,t)\sum_{m_{2}=1}^{\infty}P(m_{2},t){\sum_{m_{1}=1}^{m_{2}}}
			g_{m_{2}}(m_{1})          
                  +\sum_{m_{1}=1}^{\infty}P(m_{1},t)g_{m_{1}}(m_{1}).
\end{eqnarray}
These equations enumerate all possible ways in which a site with 
mass $m$ may change its mass. The first term on the right-hand-side
(RHS) of Eq.~(\ref{rate1}) is the
``loss'' of mass $m$ due to chipping, i.e., a site with mass $m$ may lose a fragment of mass 
$m_{1}$ ($\leq m$) to a neighbor. The second term on the RHS represents the loss due to transfer of 
mass from a neighbor chipping. The third and fourth terms are the ``gain'' terms which 
represent the ways in which a site with mass greater (lesser) than $m$ can lose (gain) 
the excess (deficit) to yield mass $m$. The terms of Eq.~(\ref{rate10}) can be interpreted 
similarly. In order to ensure that all loss and gain terms have been included 
in the rate equations, it is useful to check the sum rule 
\begin{equation}
\label{sumrule}
\frac{d}{dt}\sum_{m=0}^{\infty}P(m,t)=0, \quad \mbox{or} \quad \sum_{m=0}^{\infty}P(m,t) = 1. 
\end{equation}
With some algebra, it can be shown that Eqs.~(\ref{rate1})-(\ref{rate10}) do indeed satisfy the 
above rule. We provide the steps for this check in Appendix A.

If other microscopic processes are present, additional terms will have to be
included in the rate equations (\ref{rate1}) and (\ref{rate10}). For example, if
adsorption of a unit mass at a site occurs with rate $q$, we require additional  
terms $-qP(m,t)$ and $+qP(m-1,t)$ in Eq.~(\ref{rate1}), and $-qP(0,t)$ in Eq.~(\ref{rate10}).
While it is simple to write realistic rate equations by including all relevant
microscopic processes, it is often difficult to solve them 
analytically to obtain either time-dependent or steady-state solutions.

Several MF models studied earlier may be obtained as special cases of  
Eqs.~(\ref{rate1})-(\ref{rate10}) by an appropriate choice of the chipping 
kernel $g_{m}(n)$. Some interesting issues which have been addressed in these studies, 
in addition to obtaining steady-state mass distributions, are the 
possibilities of phase transitions in these models. We mention two representative examples to highlight the typical questions which are addressed in this area. \\
1) Majumdar et al. \cite{sk3} studied a conserved-mass model in which either
a single unit or the entire mass could dissociate from a site. Thus, $g_{m}(n)$ has the form
\begin{equation}
\label{ssm}
g_m(n) = w\delta_{n,1} + \delta_{n,m} ,
\end{equation}
where $w$ is the relative rate of the 1-chip process.
The corresponding rate equations are obtained by substituting Eq.~(\ref{ssm}) in Eqs.~(\ref{rate1})-(\ref{rate10}) as follows:
\begin{eqnarray}
\label{snm1}
\frac{d}{dt}P(m,t)&=&-(1+w)(1+s_{1})P(m,t)+wP(m+1,t)+ws_{1}P(m-1,t)\nonumber\\
&&+\sum_{m_{1}=1}^{m}P(m-m_{1},t)P(m_{1},t),\hspace{0.1in} m\geq 1, \\
\label{snm0}
\frac{d}{dt}P(0,t)&=&-(1+w)s_{1}P(0,t)+wP(1,t)+s_{1},
\end{eqnarray}
where $s_{1}=\sum_{m=1}^{\infty}P(m,t)$.

The steady-state mass distributions [$P(m)$ vs. $m$] for Eqs.~(\ref{snm1})-(\ref{snm0}) were calculated by Majumdar et al. as a function of the density $\rho = \langle m \rangle$ of the system. The relevant analytical techniques are described in Sec.~\ref{s3}. They observed a {\it dynamical phase transition} as $\rho$ was varied ($w$ being fixed), with the different phases being characterized by different steady-state distributions. 
For $\rho < \rho_{c}(w)$, $P(m)$ decayed exponentially for large $m$. For $\rho = \rho_{c}(w)$, $P(m)$ showed a power-law decay, $P(m) \sim m^{-\tau}$ with a universal exponent $\tau =5/2$. Finally, the ``high-density" phase arising for $\rho>\rho_{c}(w)$  was characterized by the formation of an infinite aggregate (at $m=\infty$). The aggregate coexisted with smaller clusters, and their mass distribution showed a power-law decay, $P(m) \sim m^{-\tau}$. \\
2) Rajesh et al. \cite{rajesh} studied a system of fragmenting and coagulating 
particles with mass-dependent diffusion rates. In this model, 
\begin{equation}
\label{rajesh2002}    
g_m(n) = w\delta_{n,1} + {m}^{-\alpha}\delta_{n,m}.  
\end{equation}
The case $\alpha$ = 0 corresponds to the model in Eq.~(\ref{ssm}). The corresponding rate equations are
\begin{eqnarray}
\label{Rajeshll}
\frac{d}{dt}P(m,t)&=&-\left(w+m^{-\alpha}+ws_{1}+\overline{s}_{1}\right)P(m,t)+wP(m+1,t)+ws_{1}P(m-1,t)\nonumber\\
&&+\sum_{m_{1}=1}^{m}\frac{P(m-m_{1},t)P(m_{1},t)}{m_{1}^{\alpha}},\hspace{0.1in} m\geq 1,\\
\label{Rajeshl0}
\frac{d}{dt}P(0,t)&=&-\left(ws_{1}+\overline{s}_{1}\right)P(0,t)+wP(1,t)+\overline{s}_{1},
\end{eqnarray} 
where $\overline{s}_{1}=\sum_{m=1}^{\infty} m^{-\alpha} P(m,t)$.

For $\alpha >0$, Rajesh et al. showed that there is no dynamical phase transition. The high-density
phase with an infinite aggregate disappears, although its imprint in the form of a
large aggregate is observed in finite systems. Further, the steady-state probability
distribution $P(m)$ decays exponentially with $m$ for all $\rho$ and $\alpha>0$.

Before concluding this discussion, a few words regarding the condensation transition
are in order. The condensation observed in the model in Eqs.~(\ref{snm1})-(\ref{snm0}) occurs due to the dynamical rules of evolution and not due to an ``attraction'' between the masses. Though it  
shares analogies with {\it Bose-Einstein condensation} (BEC), an important difference is that these condensates are formed in real space rather than momentum space as is the case in BEC. As a matter of fact, condensation occurs in a variety of seemingly diverse systems which are governed by nonequilibrium dynamics \cite{sk3}.

\subsection{Some Applications of Mass-Transport Models}

We now discuss some physical applications of mass-transport models. The aim here is 
to stress the general nature of the questions addressed in a variety of physical 
situations involving mass transport.

\subsubsection{Magnetic Nanoparticles}

Recently, there has been much research interest in  suspensions of single-domain {\it magnetic nanoparticles} (MNP), which have a wide range of technological applications, e.g., memory devices, magnetic resonance imaging, targeted drug delivery, bio-markers and bio-sensors \cite{oden,pankh}. A major reason for the utility of MNPs is the ease with which they can be detected and manipulated by an external magnetic field. Their response times are strongly size-dependent, thus introducing the possibility of controlling particle sizes to obtain desired response times.

An inherent property of MNP suspensions is cluster formation, due to the presence
of attractive interactions of varying strengths between the 
constituent particles \cite{rosen}.  Therefore, mass-transport models with fragmentation and aggregation have been traditionally employed to study clustering dynamics in these systems. 
The steady-state cluster-size distributions and the average cluster size are determined
by the interplay between aggregation (due to attractive interactions) and 
fragmentation (due to repulsive interactions and thermal noise) 
\cite{rosen}. Assuming that the number of particles is $N$, 
and denoting the number of clusters containing $k$ particles at time $t$ by $c(k,t)$, the  
rate equations in the MF approximation are as follows \cite{krap}:
\begin{eqnarray}
\label{reqn}
\frac{\partial}{\partial t}c(k,t)&=& \frac{1}{2}\sum_{i,j=1}^{\infty}\delta_{k,i+j}K_{ij}c(i,t)c(j,t) 
- c(k,t)\sum^{\infty}_{j=1} K_{kj}c(j,t) \nonumber\\
 &&+ f_{k+1}c(k+1,t) -f_{k}c(k,t) +\delta_{k,1}\sum^{\infty}_{j=1} f_{j}c(j,t),\ \ \ k \ge 1.
 \end{eqnarray}
In Eq.~(\ref{reqn}), $K_{ij}$ and $f_{k}$ are the aggregation and fragmentation kernels, 
respectively. The aggregation kernel describes the coalescence of two clusters containing
$i$ and $j$ particles to yield a larger cluster with $k$ $=$ $i+j$ particles. In many
models, it is assumed to have a mass-dependent form, $K_{ij}$ $=$ $D(i^{-\mu}+j^{-\mu})$. 
This accounts for the reduced mobility of large clusters. The fragmentation kernel $f_{k}$ describes
the loss of one particle from a cluster with $k$ particles, and has also been assumed to 
have a mass-dependent form, $f_{k} = w k^\nu $. Equation~(\ref{reqn}) can be rewritten in terms of probability distributions [cf. Eqs.~(\ref{rate1})-(\ref{rate10})] by introducing a normalization factor, $P(k,t)=c(k,t)/\sum_{k=1}^{\infty}c(k,t)$.

\subsubsection{Traffic Models}

Our second example is in the context of traffic models. In this context, we discuss the so-called {\it Bus Route Model} (BRM) \cite{loan}. Here, one is interested in the initial conditions or parameters which result in a clustering of buses or a traffic jam. The model is defined on a  
1-dimensional lattice of size $L$. Each site $i$ has two associated variables $\tau_{i}$ and $\phi_{i}$: (i) If a site $i$ is occupied by a bus, $\tau_{i}$ $=$ 1; otherwise $\tau_{i}$ $=$ 0. (ii) If a site $i$ has passengers, then $\phi_{i}$ $=$ 1; otherwise  $\phi_{i}$ $=$ 0. A site cannot have both  $\tau_{i}$ $=$ $\phi_{i}$ $=$ 1, i.e., $\tau_{i}+\phi_{i}\leq 1$.
If there are $M$ buses, the bus density $\rho$ $=$ $M/L$ is a conserved quantity.  However,
the total number of sites with passengers is not conserved.

The update rules are as follows: (i) Pick a site $i$ at random. (ii) If 
$\tau_{i}$ $=$ $\phi_{i}=$ 0, then set $\phi_{i}$ $=$ 1 with rate $\lambda$, i.e., a passenger 
arrives at an empty site with rate $\lambda$. (iii) If $\tau_{i}$ $=$ 1 and $\tau_{i+1}$ $=$ 0,
a bus hops onto a site with no passengers ($\phi_{i+1}=$ 0) with rate $\alpha$, and to a site 
with passengers ($\phi_{i+1}$ $=$ 1) with rate $\beta$. Thus, the variables $\tau_{i}\rightarrow$ 0 
and $\tau_{i+1}\rightarrow$ 1 and $\phi_{i+1}\rightarrow$ 0 with rate $\alpha$ or $\beta$, as the 
case may be. Usually, $\beta<\alpha$ as the buses slow down when passengers are being 
picked up. A jam in the system is a gap between buses of size $x\sim O(L)$, which is stable in the thermodynamic limit. 

The MF approximation of this model considers the distribution of gaps $P(x,t)$, ignoring the time-correlations in the hopping of buses. It should be noted here that, unlike the 
mass-transport models described in Sec.~\ref{s2a}, the BRM is asymmetric. Thus, the movement of the buses is unidirectional although the hop rate is proportional to the size of the gap. These features put the BRM in a class of models which are referred to as {\it zero-range processes} (ZRP) \cite{evans}. The important property of a ZRP is that it yields a steady-state as a product of marginals calculated using well-defined procedures \cite{emz04}. Further, MF calculations are exact for this class of models.

From simulations of the discrete model, heuristic arguments and MF theory, O'Loan et al. obtain evidence of a jamming transition as a function of the density of buses $\rho$. In terms of buses and passengers, the jam may be interpreted as follows. An ideal situation is one where the buses are evenly distributed along the route so that each bus picks up approximately the same number of passengers. Jamming or clustering of buses may occur if one of the buses gets delayed due to some
fluctuation at a pick-up point. Subsequently, the buses which follow catch up with the delayed bus resulting in a jam! An important observation here is that the jamming is a consequence of a local, stochastic dynamics which couples the conserved variable (buses) and the nonconserved variable (passengers). The transition is reminiscent of the condensation transition described earlier \cite{sk3}, and has also been useful in describing {\it clogging} in the transport of sticky particles down a pipe \cite{loan}.

\subsubsection{Granular Packing}

As a final example, we consider the packing of granular materials, which is 
important in many technological processes. The crucial issue in these problems is understanding
the complex network of 
forces which is responsible for the static structure and properties of granular materials. 
One such system which has been subjected to experiments, simulations and analysis is a 
pack of spherical beads in a compression cell \cite{beads}. 

The bead pack is modeled as a regular lattice of sites, each having a particle of unit mass. The
mechanisms which lead to the formation of force chains in the system are summarized in
the rules defined below:  
(i) Each site $i$ in layer $D$ is connected to $N$ sites $j$ in layer $D+1$.
(ii) Only vertical forces are considered explicitly. A fraction $q_{ij}$ of the total weight 
supported by particle $i$ in layer $D$ is transmitted to particle $j$ in layer $D+1$.
Thus, the weight $W(D,i)$ supported by the particle at the $i^{\rm th}$ site in layer $D$ satisfies the 
stochastic equation
\begin{equation}
\label{bp}
W(D+1,j) = 1 + \sum_{i} q_{ij}(D) W(D,i).
\end{equation}
The $q_{ij}(D)$ are independently-distributed random variables which satisfy the constraint
$\sum_{j}q_{ij}$ $=$ 1, required for enforcing the force-balance condition on each particle.

In general, the values of $W$ at neighboring sites in layer $D$ are not independent. The MF 
approximation of this model ignores these correlations. Defining a normalized weight variable 
$v=W/D$, we want to obtain the force distribution $P_{D}(v)$, i.e., the probability 
that a site at depth $D$ is subject to a vertical force $v$. Within the MF approximation,
it is possible to obtain a recursive equation for $P_{D}(v)$. Coppersmith et al. \cite{beads} found that, for almost all distributions of $q$, the distribution of forces decays exponentially. However,
a power-law decay was also observed in some cases.

\section{Fragmentation and Aggregation of $k$-chips}
\label{s3}

\subsection{1-chip model}
\label{s3.1}
Let us first consider the 1-chip model. The chipping kernel has the simple form
\begin{eqnarray}
\label{kernel1}
g_{m}(n)&=& w\delta_{n,1}.
\end{eqnarray}
With the above kernel, Eqs.~(\ref{rate1})-(\ref{rate10}) become
\begin{eqnarray}
\frac{d}{dt}P(m,t)&=&-P(m,t)\sum_{m_{1}=1}^m w\delta_{m_{1},1}
-P(m,t){\sum_{m_{2}=1}^{\infty}P(m_{2},t)}{\sum_{m_{1}=1}^{m_{2}}}w\delta_{m_{1},1}\nonumber\\ 
&&+{\sum_{m_{1}=1}^{\infty}P(m+m_{1},t)}w\delta_{m_{1},1}
+\sum_{m_{1}=1}^{m}P(m-m_{1},t)\sum_{m_{2}=m_{1}}^{\infty}P(m_{2},t)w\delta_{m_{1},1}, \quad m \geq 1, \nonumber \\
\ \\
\frac{d}{dt}P(0,t)&=&-P(0,t)\sum_{m_{2}=1}^{\infty}P(m_{2},t)\sum_{m_{1}=1}^{m_{2}}w\delta_{m_1,1}
+\sum_{m_{1}=1}^{\infty}P(m_{1},t)w\delta_{m_{1},1}.
\end{eqnarray}
Absorbing  $w$ into the definition of time, these equations simplify to the following form:
\begin{eqnarray}
\label{rate3}
\frac{d}{dt}P(m,t) &=&-(1+s_{1})P(m,t)+P(m+1,t)+s_{1}P(m-1,t), \ \ \ m\geq 1,\\
\label{rate30}
\frac{d}{dt}P(0,t)&=&-s_{1}P(0,t)+P(1,t).
\end{eqnarray}
Here, we have defined $s_{1}(t)=\sum_{m=1}^{\infty}P(m,t)$ as the probability of occupancy of a site with 
mass $m \geq 1$. Consequently, the probability of a site being empty is $ P(0,t)=1-s_{1}(t)$.

The above rate equations were obtained earlier in Refs.~\cite{sk3,rajesh} 
and were solved exactly. We recall this calculation 
to illustrate the {\it generating-function approach} for obtaining steady-state
solutions of such rate equations. Defining the generating function $Q(z,t)$ $=$ 
$\sum_{m=1}^{\infty}z^{m}P(m,t)$, an equation for $\partial Q/\partial t$ can be 
obtained from Eq.~(\ref{rate3}) by multiplying both sides by $z^{m}$ and summing 
over $m$:
\begin{eqnarray}
\label{gf1}
\frac{\partial}{\partial t} Q(z,t) &=& \frac{\partial }{\partial t}\sum_{m=1}^{\infty}z^{m}P(m,t)\nonumber\\
 &=&-(1+s_{1})\sum_{m=1}^{\infty}z^{m}P(m,t)+\sum_{m=1}^{\infty}z^{m}P(m+1,t)
+s_{1}\sum_{m=1}^{\infty}z^{m}P(m-1,t)\nonumber\\
&=&-(1+s_{1})Q+\frac{1}{z}\sum_{m=2}^{\infty}z^{m}P(m,t)+s_{1}z\sum_{m=0}^{\infty}z^{m}P(m,t) \nonumber\\
\label{gf11}
&=&-(1+s_{1})Q+\frac{1}{z}\left[Q-zP(1,t)\right]+s_{1}z\left[Q+P(0,t)\right].
\end{eqnarray}
Setting $\partial Q/\partial t=0$, and substituting $P(1)=s_{1}(1-s_{1})$ from the 
steady-state version of Eq.~(\ref{rate30}), we obtain
\begin{eqnarray}
\label{gen1}
Q(z)&=&\frac{s_{1}(1-s_{1})z}{(1-s_{1}z)}.
\end{eqnarray}
The value of $s_{1}$ is fixed by mass conservation, which requires that 
$\sum_{m=1}^{\infty} mP(m)$ = $\rho$, where $\rho$ is the mass density.
Putting $dQ/dz\big|_{z=1}$ = $\rho$,
we obtain
\begin{equation}
\label{rho1}
\rho=\frac{s_{1}}{1-s_{1}} \quad \mbox{or} \quad s_{1}=\frac{\rho}{1+\rho}.
\end{equation}

The steady-state distribution $P(m)$ is the coefficient of $z^{m}$ in $Q(z)$,
and can be obtained by Taylor-expanding $Q(z)$ about $z=0$. This yields 
\begin{eqnarray}
P(m)=(1-s_{1})s_{1}^{m}, \quad m \geq 1 .
\end{eqnarray}
For a more complicated function $Q(z)$, we can obtain $P(m)$ by inverting $Q(z)$. It is useful to illustrate this for the simple form of $Q(z)$ in Eq.~(\ref{gen1}). Thus,
\begin{eqnarray}
\label{steady1}
P(m)&=&\frac{1}{2\pi i}\int_{C}dz~\frac{Q(z)}{z^{m+1}}, \hspace{0.1in} m\geq 1.
\end{eqnarray}
Here, the closed contour $C$ encircles the origin in the complex plane counter-clockwise and lies inside the circle $\mid z\mid = 1/s_1$. The integral is calculated using the residue theorem.
Only those singular points which lie within $C$ (viz., $z=0$, which is a pole of order $m$) contribute to this evaluation. The associated residue is
\begin{eqnarray}
\label {res1}
\text{Res}\; f_{1}(z=0)&=&\frac{1}{(m-1)!}\frac{d^{m-1}}{dz^{m-1}}\left(\frac{Q(z)}{z}\right)\Bigg |_{z=0}\\
\label{res1c}
 &=&(1-s_{1})s_{1}^{m}.
\end{eqnarray}
Thus, the steady-state mass distribution is
\begin{eqnarray}
\label{soln1}
P(m)&=&\frac{1}{2\pi i} \cdot 2\pi i\, \text{Res}\, f_{1}(0) = (1-s_{1})s_{1}^{m}, \quad m\geq 1.
\end{eqnarray}
Notice that $P(0)=1-s_{1}$, so Eq.~(\ref{soln1}) is also valid for $P(0)$.

Using Eq.~(\ref{rho1}), the above mass distribution can be rewritten as
\begin{eqnarray}
\label{soln11}
P(m)&=&\frac{1}{1+\rho}\left( \frac{\rho}{1+\rho}\right)^{m}
\nonumber \\
&\equiv& ae^{-bm} ,
\end{eqnarray}
where 
\begin{equation}
\label{a1}
a=\frac{1}{1+\rho}, \quad b=\ln\left(\frac{1+\rho}{\rho}\right).
\end{equation}

In the case of simple chipping kernels, as in Eq.~(\ref{kernel1}), the above solution can also be obtained directly from the difference equations (\ref{rate3})-(\ref{rate30}) by setting the left-hand-side (LHS) to zero. We can then write down expressions for the first few terms of $P(m)$:
\begin{eqnarray}
\label{recur1}
P(1)&=&s_{1}P(0)=s_{1}-s_{1}^{2},\nonumber\\    
P(2)&=&(1+s_{1})P(1)-s_{1}P(0)\nonumber\\
     &=&s_{1}^{2}P(0)=s_{1}^{2}-s_{1}^{3},\nonumber\\
    P(3)&=&(1+s_{1})P(2)-s_{1}P(1)\nonumber\\
     &=&s_{1}^{3}P(0)=s_{1}^{3}-s_{1}^{4},\nonumber\\  
&.& \nonumber \\
\label{ss1de}
P(m)&=&(1+s_{1})P(m-1)-s_{1}P(m-2) \nonumber\\
&=&s_{1}^{m}P(0)=s_{1}^{m}-s_{1}^{m+1},     
\end{eqnarray}
which is identical to Eq.~(\ref{soln1}). Again, the mass conservation condition 
$\sum_{m=1}^{\infty} mP(m)$ = $\rho$ results in Eq.~(\ref{rho1}), as expected.

The 1-chip solution in Eq.~(\ref{soln11}) is important because of its universal
nature. As a matter of fact, it is a steady-state solution for all MF
models where the chipping kernel $g_{m}(n)$ is independent of the mass of
the departure site $m$, $g_m(n)=g(n)$. To confirm this, we consider Eqs.~(\ref{rate1})-(\ref{rate10}) with $g_{m}(n)$ replaced by $g(n)$. The corresponding rate equations are
\begin{eqnarray}
\label{arbk1}
\frac{d}{dt}P(m,t)&=&-P(m,t)\sum_{m_{1}=1}^m g(m_{1})
      -P(m,t)\sum_{m_{2}=1}^{\infty}P(m_{2},t)\sum_{m_{1}=1}^{m_{2}}
	g(m_{1})\nonumber \\
  & &+\sum_{m_{1}=1}^{\infty}P(m+m_{1},t)g(m_{1})\nonumber\\
      & & +\sum_{m_{1}=1}^{m}P(m-m_{1},t)\sum_{m_{2}=m_{1}}^{\infty} 
    P(m_{2},t)g(m_{1}), \ \ \ m \ge 1,
\end{eqnarray}
\begin{eqnarray}
\label{arbk10}
\frac{d}{dt}P(0,t)&=&-P(0,t)\sum_{m_{2}=1}^{\infty}P(m_{2},t){\sum_{m_{1}=1}^{m_{2}}}
			g(m_{1})          
                  +\sum_{m_{1}=1}^{\infty}P(m_{1},t)g(m_{1}).
\end{eqnarray}

In the steady state, the above equations may be combined to obtain 
\begin{eqnarray}
\label{arbk1ss}
&&-P(m)\sum_{m_{1}=1}^{m} g(m_{1})-P(m)\frac{1}{P(0)}
		\sum_{m_{1}=1}^{\infty}P(m_{1})g(m_{1})\nonumber\\
&&      +\sum_{m_{1}=1}^{\infty}P(m+m_{1})g(m_{1})
+\sum_{m_{1}=1}^{\infty}P(m-m_{1})\sum_{m_{2}=m_{1}}^{\infty}P(m_{2}) g(m_{1})=0. 
\end{eqnarray}
Substituting $P(m)=a \exp(-bm)$ on the RHS of Eq.~(\ref{arbk1ss}), we obtain
\begin{eqnarray}
\text{RHS} &=& \sum_{m_{1}=1}^{m} g(m_{1})-\frac{1}{a}\sum_{m_{1}=1}^{\infty}ae^{-bm_{1}}g(m_{1}) +\sum_{m_{1}=1}^{\infty} e^{-bm_{1}}g(m_{1}) \nonumber\\
&& +\sum_{m_{1}=1}^{\infty}\sum_{m_{2}=m_{1}}^{\infty}ae^{-b\left(m_{2}-m_{1}\right)} g(m_{1}) ,
\end{eqnarray}
The first and fourth terms cancel, and the second and third terms cancel, so $\text{RHS} = 0$. This confirms that $P(m)=a e^{-bm}$ is a solution of Eqs.~(\ref{arbk1})-(\ref{arbk10}). The constants $a$ and $b$ are fixed by the requirements of probability normalization $[\sum_{m=0}^{\infty}P(m) = 1]$ and mass conservation [$\sum_{m=1}^{\infty}mP(m) = \rho$]. We leave it as an exercise to the reader
to verify that these conditions lead to the same values of $a$ and $b$ as in Eq.~(\ref{a1}).

Next, let us generalize the $1$-chip model to a $k$-chip model, where $k>1$. These
$k$-chip models are interesting in physical situations where the deposited
material consists of polymers or aggregates. We will see that the steady-state
solutions for $k$-chip models exhibit a $k$-branch structure.

\subsection{2-chip model}
\label{s3.2}

The steady-state distributions for the 2-chip model can be
obtained using a procedure similar to the 1-chip model. The corresponding
form of $g_m(n)$ is
\begin{equation}
\label{kernel2}
g_{m}(n)=w\delta_{n,2}.
\end{equation}
The rate equations in this case are (absorbing $w$ into time $t$)
\begin{eqnarray}
\label{rate2}
\frac{d}{dt}P(m,t)&=&-(1+s_{2})P(m,t)+P(m+2,t)+s_{2}P(m-2,t), \quad m \ge 2, \\
\label{rate20}
\frac{d}{dt}P(m,t)&=&-s_{2}P(m,t)+P(m+2,t), \quad m < 2.
\end{eqnarray}
Here, $s_{2}(t)$ = $\sum_{m=2}^{\infty}P(m,t)$ is the probability of sites having mass 2 or more.
Notice that the kernel in Eq.~(\ref{kernel2}) is independent of the mass of the departure site.
Thus, the 1-chip solution is a steady-state  solution of Eqs.~(\ref{rate2})-(\ref{rate20}),
as can be verified by direct substitution. However, an arbitrary initial condition $P(m,0)$ will
not relax to this solution due to the presence of conserved quantities, as we shall see shortly.

The steady-state generating function $Q(z)$, obtained in analogy with the $1$-chip case, is
as follows:
\begin{equation}
\label{gen2}
 Q(z)=\frac{z(s_{1}-s_{2})+z^{2}s_{2}(1-s_{1})}{(1-s_{2}z^{2})},
\end{equation}
where $s_{2}$ = $\sum_{m=2}^{\infty}P(m)$.
It is straightforward to Taylor-expand $Q(z)$ in Eq.~(\ref{gen2}) and identify $P(m)$. Alternatively, we can obtain $P(m)$ using Eq.~(\ref{steady1}). Thus
\begin{eqnarray}
P(m)&=&\frac{1}{2\pi i}\int_{C}dz\frac{(s_{1}-s_{2})
        +s_{2}(1-s_{1})z}{z^{m} \left( 1-s_{2}z^{2}\right)}, \quad m\geq 1.
\end{eqnarray}
In this case, the contour $C$ encircles the origin counterclockwise, and lies inside the circle 
$\mid z\mid$ = $1/\sqrt{s_{2}}$. The singularities of the integrand $f_2(z)$ in the above equation are $z=0$ (pole of order $m$),
$z=1/\sqrt{s_{2}}$ (simple pole) and $z$ = $-1/\sqrt{s_{2}}$ (simple pole). The second
and third poles lie outside $C$, making Res~$f_{2}(z=0)$ the only contributing residue.  
This evaluation yields
\begin{eqnarray}
\label{res2}
\text{Res}~f_{2}(z=0)&=&\frac{(1-s_{1})}{2}\left[1+(-1)^{m}\right]s_{2}^{m/2}
+\frac{(s_{1}-s_{2})}{2}\left[1-(-1)^{m}\right]s_{2}^{(m-1)/2}.
\end{eqnarray}
Thus, the steady-state probability distribution for the 2-chip model is given by
\begin{eqnarray}
\label{steady2}
P(m)&=& (1-s_{1})s_{2}^{m/2}\delta_{\mbox{mod}(m,2),0} 
+ (s_{1}-s_{2})s_{2}^{(m-1)/2}\delta_{\mbox{mod}(m,2),1} \nonumber \\
\label{steady22}
&\equiv& P^{\text{e}}(m) + P^{\text{o}}(m) .
\end{eqnarray}
Here the function $\mbox{mod}(m,n)$ is defined as the remainder on division of $m$ by $n$.
The first and second terms on the RHS of Eq.~(\ref{steady2}) are the steady-state
distributions for even values of $m$ $\left[P^{\text{e}}(m)\right]$ and odd values
of $m$ $\left[P^{\text{o}}(m)\right]$, respectively. Thus, the 2-chip model has a steady-state
solution comprising of two branches, both of which have the same exponential decay.
Notice that the occupation probabilities for sites with even or odd units of mass are
\begin{eqnarray}
\label{seso} 
S_{\text{e}} &=& \sum_{m=0,2,4,..}^{\infty}P^{\text{e}}(m) = \frac{1-s_{1}}{1-s_{2}}, \\
S_{\text{o}} &=& \sum_{m=1,3,5,..}^{\infty}P^{\text{o}}(m) = \frac{s_{1}-s_{2}}{1-s_{2}}.
\end{eqnarray}
These quantities remain conserved during the evolution because of the nature of the
2-chip move. The two branches appear as a consequence of these two conserved quantities.

The probabilities of occupancy $s_{1}$ and $s_{2}$ are related to the mass
density $\rho$ as follows:
\begin{equation}
\label{rho2}
\rho= \frac{dQ}{dz} \bigg|_{z=1} = \frac{s_{1}+s_{2}}{1-s_{2}} .
\end{equation}
As before, $\rho$  may be calculated from either $Q(z)$ or $P(m)$.
The quantities $s_{1}$ and $s_{2}$  can be determined in terms of $\rho$ and $S_{\text{e}}$ (or $S_{\text{o}}$):
\begin{equation}
\label{s1}
s_{1} = \frac{\rho - S_{\text{e}} + 1}{\rho + S_{\text{e}} + 1}, \quad s_{2} = \frac{\rho + S_{\text{e}} -1}
{\rho + S_{\text{e}} + 1}.
\end{equation}
It should be noted that $\rho + S_{\text{e}}$ is always greater than 1, ensuring $s_{2}> 0$.
If we choose the initial conditions $P(m,0)$ such that $s_{2}=s_{1}^{2}$ in Eq.~(\ref{s1}),
the branched solution of Eq.~(\ref{steady2}) reduces to the 1-chip exponential solution.
Alternatively, if we substitute $s_{2}=s_{1}^{2}$ in Eq.~(\ref{gen2}), we recover
the generating function of the 1-chip model in Eq.~(\ref{gen1}).

\subsection{$k$-chip model}
\label{s3.3}

In general, consider the case of $k$ units of mass chipping from a site and
then aggregating with the mass of a randomly-chosen nearest neighbor: 
\begin{equation}
\label{kernelk}
g_{m}(n)=w\delta_{n,k}.
\end{equation}
The corresponding rate equations for $P(m,t)$, obtained by substituting
Eq.~(\ref{kernelk}) in Eqs.~(\ref{rate1})-(\ref{rate10}), are as follows
(absorbing $w$ into $t$):
\begin{eqnarray}
\label{ratek}
\frac{d}{dt}P(m,t)&=&-(1+s_{k})P(m,t)+P(m+k,t)+s_{k}P(m-k,t), \quad m\geq k,\\
\label{ratek0}
\frac{d}{dt}P(m,t)&=&-s_{k}P(m,t)+P(m+k,t), \quad m < k .
\end{eqnarray}
Here, $s_{k}(t)$ = $\sum_{m=k}^{\infty}P(m,t)$ is the probability of sites having
mass $k$ or more. As the kernel in Eq.~(\ref{kernelk}) is independent
of the mass of the departure site, $P(m)=a \exp(-bm)$ is a steady-state
solution of Eqs.~(\ref{ratek})-(\ref{ratek0}). However, as in the 2-chip case, an arbitrary initial condition $P(m,0)$ will not relax to this exponential solution due to the presence of $k$ conserved quantities: $\sum_{m=0}^{\infty}P(n+mk,t)$ with $n$ $=$ $0,1,...,k-1.$ Rather, the 
steady-state solution will consist of $k$ branches. 

The steady-state generating function for the $k$-chip model is 
\begin{equation}
\label{genk}
Q(z) = \frac{z(s_{1}-s_{2})+z^{2}(s_{2}-s_{3})+\cdot\cdot\cdot\cdot+z^{k}s_{k}(1-s_{1})}
	{(1-s_{k}z^{k})},
\end{equation}
where $s_{k}$ $=$ $\sum_{m=k}^{\infty}P(m)$. The corresponding probability distribution
is as follows:
\begin{eqnarray}
\label{steadym}
P(m)&=&(1-s_{1})s_k^{m/k}\delta_{\mbox{mod}(m,k),0}
		+(s_{1}-s_{2})s_k^{(m-1)/k}\delta_{\mbox{mod}(m,k),1}+....+ \nonumber\\
	& &(s_{i}-s_{i+1})s_k^{(m-i)/k}\delta_{\mbox{mod}(m,k),i}+....+
	(s_{k-1}-s_{k})s_k^{(m-k+1)/k}\delta_{\mbox{mod}(m,k),k-1}.
\end{eqnarray}
For completeness, we present the derivation of Eq.~(\ref{steadym}) for the $3$-chip case in Appendix B. The $s_{i}$'s are related to the mass density via the relation 
\begin{equation}
\label{rhok}
\rho=\frac{s_{1}+s_{2}+.......+s_{k}}{1-s_{k}}.
\end{equation}
Thus, the steady-state solution for the $k$-chip model consists of $k$ branches. All
the branches in the probability distribution of Eq.~(\ref{steadym}) decay exponentially
with a slope $\ln (s_{k})/k$. The quantities $s_{1},.....,s_{k}$ are determined from Eq.~(\ref{rhok})  plus the $(k-1)$ conserved probability sums in the branches:
 \begin{eqnarray}
 \label{conserv}
 S_{i}&=&\sum_{m=i,i+k,...}^{\infty}(s_{i}-s_{i+1})s_{k}^{(m-i)/k}\nonumber\\
 &=&\frac{s_{i}-s_{i+1}}{1-s_{k}}, \hspace{0.1in} i=0,1....,k-1.
 \end{eqnarray}
 Notice that one of the $S_{i}$'s (say, $S_{k-1}$) is not independent because $\sum_{i=0}^{k-1}S_{i}=1$
 Further, appropriately chosen initial conditions resulting 
in steady-state values $s_{2}=s_{1}^{2}$, $s_{3}=s_{1}^{3}$,....,$s_{k}=s_{1}^{k}$ 
collapse the $k$ branches in Eq.~(\ref{steadym}) to the 1-chip solution.

Before concluding this subsection, let us make an observation about the $k$-chip lattice
model, i.e., the original model rather than its ``rate equation" counterpart. This model has an exponentially large set of disjoint sectors,
and configurations in different sectors are not connected by the dynamics. To see this,
we denote the number of particles on a site $i$ as $m_i$. As $k$ particles arrive at or
leave this site at a given time, the quantity $M_i=\mbox{mod}(m_i,k)$ is conserved. Therefore,
the set $\{M_i\}$ is conserved by the dynamics, and labels a particular sector. The
number of sectors is $k^N$, where $N$ is the number of lattice sites. Such systems have
been referred to as {\it many-sector decomposable systems} by Menon et al. \cite{mbd97}.

\section{Monte Carlo Simulations}
\label{s4}

In this section, we present Monte Carlo (MC) results for some of the models
discussed earlier. All simulations were performed on 1-$d$ and 2-$d$ lattices with
periodic boundary conditions. The lattice sizes were $L$ $=$ $1024$ (in $d=1$) and 
$L^{2}$ $=$ $128\times 128$ (in $d=2$). The data presented here was obtained
as an average over $500$ independent runs. The details of the MC procedure are provided in Appendix C, so that the reader can implement these models numerically.
 
First, we present results for chipping kernels which satisfy $g_{m}(n) = g(n)$,
discussed at the end of Sec.~\ref{s3.1}. In Fig.~2, we plot $P(m)$ vs. $m$ obtained from
1-$d$ MC simulations with three different functional forms of $g_{m}(n)$:
\begin{eqnarray}
g_{m}(n)&=&1,\nonumber\\
\label{power}
g_{m}(n)&=&\frac{1}{n^{2}},\nonumber\\
g_{m}(n)&=&e^{-0.1n} .
\end{eqnarray}
The MC data sets are numerically coincident with each other as well as the 1-chip solution 
in Eq.~(\ref{soln11}) (denoted as a solid line), which was obtained from the corresponding MF
equations. The mass density in each case was $\rho=5$.
There is excellent agreement between the different data sets, showing that the MC data is
described very well by the solutions of the corresponding MF equations, even for $d=1$.
This feature is also observed in our subsequent results, suggesting that the MF equations 
are exact in the present context \cite{emz04}. 

Next, we present results for the 2-chip model discussed in Sec.~\ref{s3.2}. Figure~3(a)
shows the steady-state distribution obtained from 1-$d$ and 2-$d$ MC simulations
for initial conditions with $\rho=10$, and $S_{\text{e}}=1, S_{\text{o}}=0$.
The solid line denotes the result in Eq.~(\ref{steady2}) with values of 
$s_{1}$ and $s_{2}$ evaluated from Eq.~(\ref{s1}). As our initial condition only had
sites with even $m$ populated, the steady-state solution is the even-$m$ branch of 
Eq.~(\ref{steady2}). Figure~3(b) is similar, but for a mixed initial condition 
with $\rho=9.5$ and $S_{\rm e}=S_{\rm o}=1/2$. The branched nature of the solution,
resulting in a staircase-type probability distribution, is highlighted in the inset.

In Fig.~4, we show the steady-state distributions obtained from 1-$d$ and 2-$d$ MC simulations of 
the 3-chip model. In Fig.~4(a), the initial condition had $P(9,0)=P(10,0)=P(11,0)=1/3$.
This corresponds to $\rho=10$, and all three branches are equally populated. The solid line 
in Fig.~4(a) denotes the result in Eq.~(\ref{steadym}) with $s_{1}, s_{2}$ 
and $s_{3}$ calculated from Eq.~(\ref{3chs}). Figure~4(b) is analogous to Fig.~4(a), but
the initial condition now has $P(9,0)=1/2$, $P(10,0)=1/3$, and $P(11,0)=1/6$. The
corresponding value of the average density is $\rho \simeq 9.67$. Again, in both sets
of figures, the MC simulations agree very well with the corresponding MF result.

\section{Summary and Discussion}
\label{s5}

Let us conclude this paper with a summary and discussion. There has been much research interest in mass-transport models, which arise in many physical contexts. Therefore, we believe it is appropriate to make this subject accessible to a wider audience. This is the underlying motivation for this paper. For the purposes of this exposition, we focus on fragmentation-aggregation models with conserved mass, i.e., there is no ongoing adsorption or desorption. We consider models with the
chipping rate $g_m(n)$, where $n$ is the chipped mass and $m$ is the mass of the
departure site. We use the corresponding mean-field (MF) equations to obtain the
steady-state probability distributions [$P(m)$ vs. $m$] for different functional
forms of $g_m(n)$. We show that a large class of chipping kernels, where $g_m(n)$
is independent of $m$, give rise to an exponentially-decaying distribution:
$P(m)=a\exp(-bm)$. This is also the MF solution for the 1-chip model [cf. Eq.~(\ref{soln11})],
where one unit of mass fragments from a site and aggregates with a randomly-chosen
nearest-neighbor.
 
We have also discussed $k$-chip models for fragmentation and aggregation.
The resulting steady-state distribution has 
$k$ branches, each of which decays exponentially with the same slope. This slope
is determined by the average density $\rho$, and the population of the branches in
the initial condition $P(m,0)$ for the rate equations (\ref{rate1})-(\ref{rate10}). The initial population in each of the branches is conserved during the evolution, and is also reflected in the steady-state distribution.
 
Finally, we compared the MF analytical results with those from Monte
Carlo (MC) simulations in $d =1,2$. In all cases, we found that the MF
results for $P(m)$ vs. $m$ were in excellent
agreement with the MC results. This demonstrates that the MF results are exact
in the present context.

There are many open research problems in the area of mass transport, aggregation and growth. These models can be tackled with a wide range of analytical and numerical techniques, which are both simple and elegant. We hope that this review will motivate further studies of this fascinating area.

\subsubsection*{\bf Acknowledgments}

The authors would like to thank M. Barma, S.N. Majumdar and R. Rajesh
for fruitful discussions. GPS and VB would like to acknowledge the support 
of CSIR Grant No. 03(1077)/06/EMR-II. SP is grateful to the Department of
Science and Technology, India for supporting this work through the
project {\it Pattern Formation in Granular Materials}. 

\newpage
\appendix
\section{Verification of the sum rule on $P(m,t)$}

We want to show that $d\left[\sum_{m=0}^{\infty}P(m,t)\right]/d t=0$. 
Using Eqs.~(\ref{rate1})-(\ref{rate10}),
\begin{eqnarray}
\label{sum1}
\frac{d}{dt}\left[\sum_{m=0}^{\infty}P(m,t)\right]&=&\frac{d}{dt}\left[\sum_{m=1}^{\infty}P(m,t)+P(0,t)\right]\nonumber\\
&=&-\sum_{m=1}^{\infty}P(m,t)\sum_{m_{1}=1}^{m}g_{m}(m_{1})
-\sum_{m=1}^{\infty}P(m,t)\sum_{m_{2}=1}^{\infty}P(m_{2},t)\sum_{m_{1}=1}^{m_{2}}g_{m_{2}}(m_{1})\nonumber\\
&&+\sum_{m=1}^{\infty}\sum_{m_{1}=1}^{\infty}P(m+m_{1},t)g_{m+m_{1}}(m_{1})\nonumber\\
&&+\sum_{m=1}^{\infty}\sum_{m_{1}=1}^{m}P(m-m_{1},t)\sum_{m_{2}=m_{1}}^{\infty}P(m_{2},t)g_{m_{2}}(m_{1})\nonumber\\
&&-P(0,t)\sum_{m_{2}=1}^{\infty}P(m_{2},t)\sum_{m_{1}=1}^{m_{2}}g_{m_{2}}(m_{1})
+\sum_{m_{1}=1}^{\infty}P(m_{1},t)g_{m_{1}}(m_{1}).
\end{eqnarray}
We regroup terms and write
\begin{eqnarray}
\label{sum2}
\text{LHS} &=&-\sum_{m=1}^{\infty}P(m,t)\sum_{m_{1}=1}^{m}g_{m}(m_{1})\nonumber\\
&&+\Bigg[-\sum_{m=1}^{\infty}P(m,t)\sum_{m_{2}=1}^{\infty}P(m_{2},t)\sum_{m_{1}=1}^{m_{2}}g_{m_{2}}(m_{1})
\nonumber\\
&&-P(0,t)\sum_{m_{2}=1}^{\infty}P(m_{2},t)\sum_{m_{1}=1}^{m_{2}}g_{m_{2}}(m_{1})\Bigg]\nonumber\\
&&+\left[\sum_{m=1}^{\infty}\sum_{m_{1}=1}^{\infty}P(m+m_{1},t)g_{m+m_{1}}(m_{1})
+\sum_{m_{1}=1}^{\infty}P(m_{1},t)g_{m_{1}}(m_{1})\right]\nonumber\\
&&+\sum_{m=1}^{\infty}\sum_{m_{1}=1}^{m}P(m-m_{1},t)\sum_{m_{2}=m_{1}}^{\infty}P(m_{2},t)g_{m_{2}}(m_{1})\\
\label{sum22}
&=&-\sum_{m=1}^{\infty}P(m,t)\sum_{m_{1}=1}^{m}g_{m}(m_{1})
-\sum_{m=0}^{\infty}P(m,t)\sum_{m_{2}=1}^{\infty}P(m_{2},t)\sum_{m_{1}=1}^{m_{2}}g_{m_{2}}(m_{1})\nonumber\\
&&+\sum_{m=0}^{\infty}\sum_{m_{1}=1}^{\infty}P(m+m_{1},t)g_{m+m_{1}}(m_{1})\nonumber\\
&&+\sum_{m=1}^{\infty}\sum_{m_{1}=1}^{m}P(m-m_{1},t)\sum_{m_{2}=m_{1}}^{\infty}P(m_{2},t)g_{m_{2}}(m_{1}) .
\end{eqnarray}

Our initial condition for Eqs.~(\ref{rate1})-(\ref{rate10}) satisfies $\sum_{m=0}^{\infty}P(m,0)=1$. Therefore, we set $\sum_{m=0}^{\infty}P(m,t)=1$ on the RHS of Eq.~(\ref{sum22}). This is justified subsequently as it results in $d\left[\sum_{m=0}^{\infty}P(m,t)\right]/d t=0$. Thus 
\begin{eqnarray}
\label{sum3}
\text{LHS}&=&-\sum_{m=1}^{\infty}P(m,t)\sum_{m_{1}=1}^{m}g_{m}(m_{1})
-\sum_{m_{2}=1}^{\infty}P(m_{2},t)\sum_{m_{1}=1}^{m_{2}}g_{m_{2}}(m_{1})\nonumber\\
&&+\sum_{m_{1}=1}^{\infty}\sum_{m_{2}=m_{1}}^{\infty}P(m_{2},t)g_{m_{2}}(m_{1})\nonumber\\
&&+\sum_{m_{1}=1}^{\infty}\sum_{m=m_{1}}^{\infty}P(m-m_{1},t)\sum_{m_{2}=m_{1}}^{\infty}P(m_{2},t)g_{m_{2}}(m_{1})\\
\label{sum4}
&=&-2\sum_{m=1}^{\infty}P(m,t)\sum_{m_{1}=1}^{m}g_{m}(m_{1})
+2\sum_{m_{1}=1}^{\infty}\sum_{m_{2}=m_{1}}^{\infty}P(m_{2},t)g_{m_{2}}(m_{1}).
\end{eqnarray}
In the second term on the RHS of Eq.~(\ref{sum4}), we interchange the order of the summations over $m_{1}$ and $m_{2}$. This leads to a cancellation of the two terms, proving that $d\left[\sum_{m=0}^{\infty}P(m,t)\right]/d t=0$.

\newpage
\section{Steady-State Distribution for the 3-chip Model}

The generating function for the 3-chip model is [cf. Eq.~(\ref{genk})]
\begin{eqnarray}
\label{3cqz}
Q(z)&=&\frac{(s_{1}-s_{2})z+(s_{2}-s_{3})z^{2}+s_{3}(1-s_{1})z^{3}}{(1-s_{3}z^{3})},
\end{eqnarray}
where $s_{3}=\sum_{m=3}^{\infty}P(m)$. From Eq.~(\ref{steady1}), we have
\begin{eqnarray}
P(m)&=&\frac{1}{2\pi i}\int_{C}dz~\frac{(s_{1}-s_{2})z+(s_{2}-s_{3})z^{2}+s_{3}(1-s_{1})z^{3}}
{z^{m+1} \left( 1-s_{3}z^{3}\right)}, \quad m \geq 1 .
\end{eqnarray}
The closed contour $C$ encircles the origin and lies inside the region defined by $\mid z \mid < 1/{s_3}^{1/3}$. As usual, only Res~$f_{3}(z=0)$ contributes to the integral, and is evaluated as
\begin{eqnarray}
\text{Res}\;f_{3}(z=0)&=&\frac{1}{(m-1)!}\frac{d^{m-1}}{dz^{m-1}}\left(\frac{Q(z)}{z}\right)
\Bigg|_{z=0} \nonumber\\
&=&\frac{1}{(m-1)!}\frac{d^{m-1}}{dz^{m-1}}
\left[\frac{(s_{1}-s_{2})+(s_{2}-s_{3})z+s_{3}(1-s_{1})z^2}
{(-s_{3})\left( z^{3}-1/s_{3}\right)}\right]\Bigg|_{z=0} \nonumber\\
\label{res3c}
&=&\frac{(1-s_{1})}{3}\left[1+2\cos\left(\frac{2\pi(m-3)}{3}\right)\right]s_{3}^{m/3}\nonumber\\
&&+\frac{(s_{1}-s_{2})}{3}\left[1+2\cos\left(\frac{2\pi(m-1)}{3}\right)\right]s_{3}^{(m-1)/3}
\nonumber\\
&&+\frac{(s_{2}-s_{3})}{3}\left[1+2\cos\left(\frac{2\pi(m-2)}{3}\right)\right]s_{3}^{(m-2)/3}.
\end{eqnarray}

Then, the steady-state mass distribution is
\begin{eqnarray}
P(m)&=&\frac{(1-s_{1})}{3}\left[1+2\cos\left(\frac{2\pi(m-3)}{3}\right)\right]s_{3}^{m/3}
\nonumber\\
&&+\frac{(s_{1}-s_{2})}{3}\left[1+2\cos\left(\frac{2\pi(m-1)}{3}\right)\right]s_{3}^{(m-1)/3}
\nonumber\\
&&+\frac{(s_{2}-s_{3})}{3}\left[1+2\cos\left(\frac{2\pi(m-2)}{3}\right)\right]s_{3}^{(m-2)/3}
\nonumber\\
\label{3csol3}
&=&(1-s_{1}){s_{3}}^{m/3}\delta_{{\rm mod}(m,3),0}
+(s_{1}-s_{2}){s_{3}}^{\left(m-1\right)/3}\delta_{{\rm mod}(m,3),1}\nonumber\\
&&+(s_{2}-s_{3}){s_{3}}^{\left(m-2\right)/3}\delta_{{\rm mod}(m,3),2}.
\end{eqnarray}
The relation between the mass density $\rho$ and the $s_{i}$'s can be obtained directly by using Eq.~(\ref{3csol3}). Thus
\begin{eqnarray}
\label{rho3c}
\rho=\frac{s_1+s_2+s_3}{1-s_{3}} .
\end{eqnarray}
As for the 2-chip model, the $s_{i}$'s can be obtained as a function of $\rho$ and the probability sums on each of the branches in Eq.~(\ref{conserv}). These result in the following expressions:
\begin{eqnarray}
\label{3chs}
s_{1} &=& \frac{\rho}{\rho+2+S_{0}-S_{2}} 
+ \frac{2S_{1}+S_{0}S_{2}-S_{1}S_{2}}{(\rho+2+S_{0}-S_{2})(1-S_{2})}, \nonumber \\
s_{2} &=& \frac{\rho}{\rho+2+S_{0}-S_{2}} + \frac{S_{2}-S_{1}}{\rho+2+S_{0}-S_{2}}, \nonumber \\
s_{3} &=& \frac{\rho}{\rho+2+S_{0}-S_{2}} 
- \frac{S_{1}+S_{2}+S_{0}S_{2}-S_{2}^{2}}{(\rho+2+S_{0}-S_{2})(1-S_{2})}.
\end{eqnarray}

\newpage
\section{Monte Carlo Simulations of Mass-Transport Models}

\subsection{1-Chip Model}

Consider a $1$-dimensional lattice with periodic boundary conditions and label the sites
with integers $i= 1, 2, 3,.., L$. Fix the mass density $\rho$.  
\begin{enumerate}
\item Initializing the lattice: Integer masses $m_{i}$ are placed on the lattice sites in accordance 
with the chosen 
$\rho$ such that $\sum_{i=1}^{L}m_{i} = \rho L$. A simple procedure for achieving this is 
as follows:
\begin{enumerate}
\item Choose an integer random number NRAN($i$) from the range $[0,L]$ for each site $i$.
\item Assign $m_{i} = \text{Int}\left[(\rho+1) L*\text{NRAN}(i)/\sum_{i}\text{NRAN}(i)\right]$, where $\text{Int}(x)$ refers to the integer part of $x$.
\end{enumerate}
\item Chipping and aggregation: A site $i$ is chosen at random. If $m_{i}$ is non-zero, a
unit mass chips and aggregates with a randomly-chosen neighbor. Thus $m_{i}\rightarrow m_{i}-1$, 
and $m_{i-1}\rightarrow m_{i-1}+1$ or $m_{i+1}\rightarrow m_{i+1}+1$ with a probability 1/2. 
\item Repeat step 2 for $L$ times, which corresponds to one MC step (MCS).
\item Compute the mass distribution of lattice sites.
\item Repeat steps 2-4 for several MCS, storing the mass distribution at intermediate
MCS.
\item Repeat steps 2-5 for a large number of independent lattice configurations, generated via step 1.
\item Compute the configuration-averaged steady-state mass distribution, $P(m)$ vs. $m$.
\end{enumerate}
The steady state is achieved when the mass distribution does not change (apart from numerical fluctuations) in subsequent MCS.

\subsection{Case with $g_{m}(n)=1/n^{2}$}

In general, the fragmentation rates $g_{m}(n)$ can be computed prior to the simulation and stored in 
a matrix for easy look-up using the following tips:
\begin{enumerate}
\item Define a matrix $G$ of dimension $N\times N$, chosen specific to $g_{m}(n)$. For example, if $g_m(n)=1/n^{2}$, the rate of chipping 
a mass of size 1000 is $10^{-6}$.  We can assume that chipping of masses greater than 
1000  occurs with a very small rate, and hence these events may be ignored. Thus we 
may set $N=1000$. 

\item Initialize $G_{mn}$ to zero. The row index $m$ corresponds to 
the mass of the lattice site chosen to fragment. Masses greater than $N$ may be 
treated as $N$ for the computation of fragmentation rates for reasons discussed 
in 1 above. The column index
refers to the chipped mass $n\leq m$. Thus the matrix $G$ has a triangular
form, with the non-zero entries calculated using $g_{m}(n)$. For example, if $g_m(n)=1/n^{2}$, the rates for the
$10^{\text{th}}$ row are [1, 0.25, 0.1111, 0.0625, 0.04, 0.02777, 0.0204, 0.0156, 0.0123, 
0.01, 0, 0, ..., 0]. To connect these to probabilities we normalize these numbers by $\sum_{n=1}^{\infty}1/n^{2}$.
\end{enumerate}

Once the look-up table for fragmentation rates is computed, the MC procedure 
is as before with step 2 replaced by the following steps.
\begin{enumerate}
\renewcommand{\labelenumi}{\Alph{enumi}.}
\item A site $i$ is chosen at random. If $m_{i}$ is non-zero, draw a random number $r$ in 
the interval (0,1).
\item Go to the $m_i^{\rm th}$ row of the table and check the two consecutive entries 
which sandwich $r$. The column number of the larger entry is the number of mass 
particles $n$ that chip from $m_i$. Thus $m_{i}\rightarrow m_{i}-n$,
and $m_{i-1}\rightarrow m_{i-1}+n$ or $m_{i+1}\rightarrow m_{i+1}+n$ with probability 1/2.
\end{enumerate}

\newpage

\begin {figure}[htb!]
\centering
\includegraphics[width=12.0cm,height=10cm,angle=0]{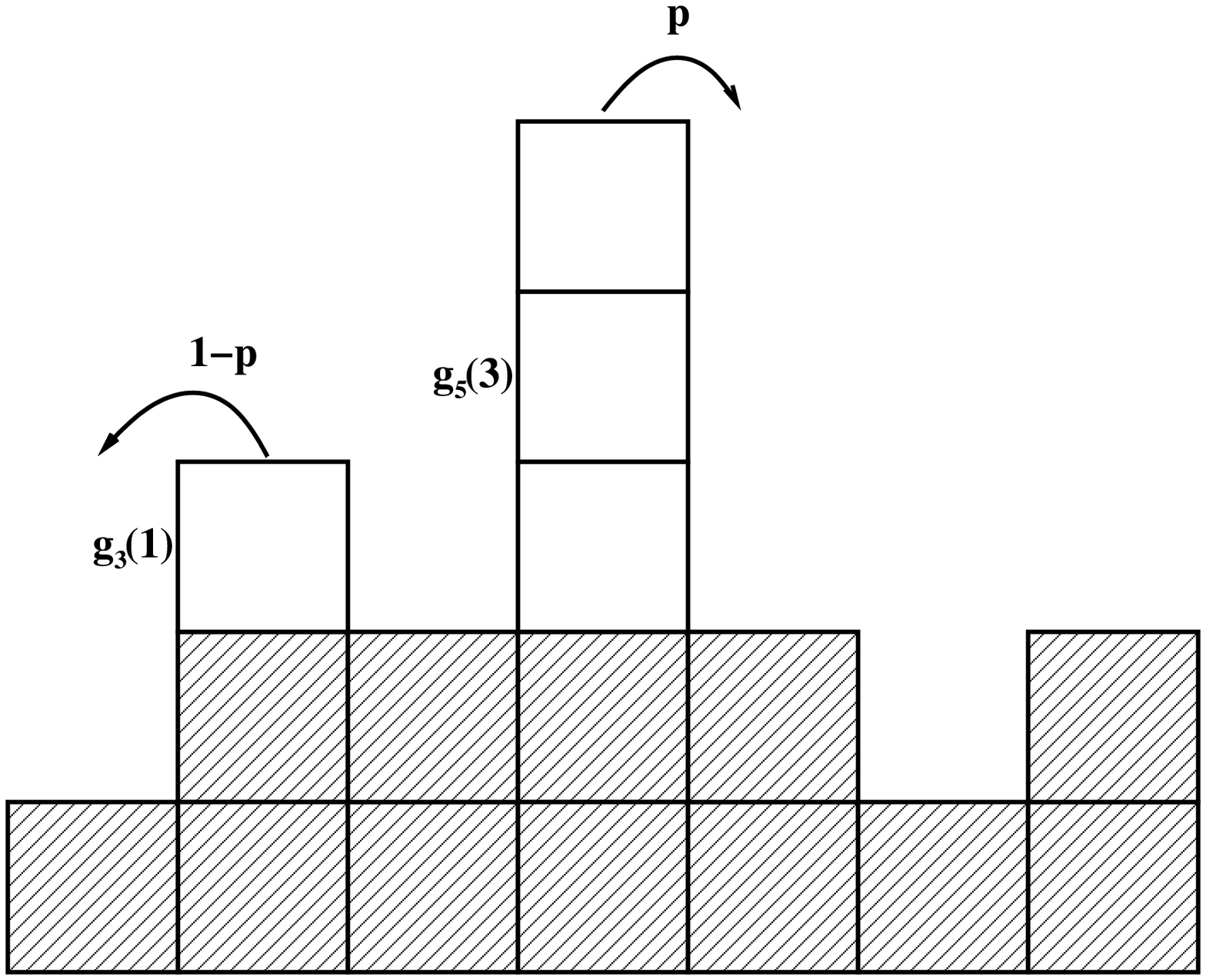}\\
\caption{Schematic representation of the conserved-mass model
with fragmentation and aggregation. A mass $n$ can chip from a site with mass $m$ 
with a fragmentation rate $g_{m}(n)$, and aggregate with the right (left) neighbor
with probability $p~(1-p)$.}
\label{Figure 1}
\end {figure}

\begin {figure}[htb!]
\centering
\includegraphics[width=12.0cm,height=10cm,angle=0]{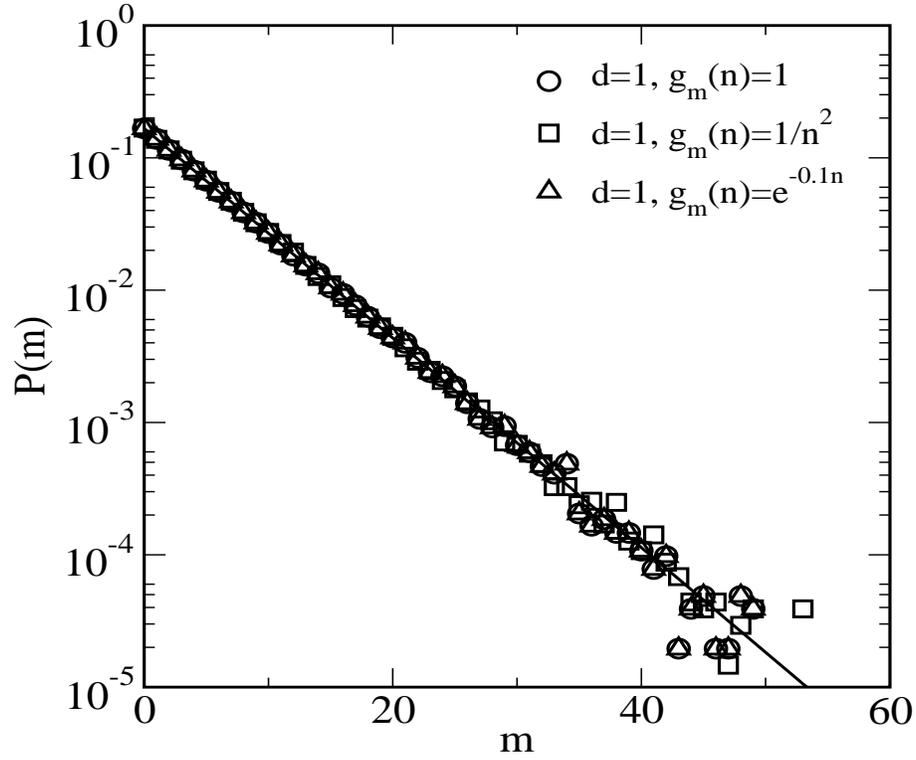}\\
\caption{Steady-state probability distributions [$P(m)$ vs. $m$] from
1-$d$ MC simulations with three different functional forms for the chipping kernel
$g_m(n)$. The data sets are plotted on a linear-logarithmic scale. The details of the
MC simulations are provided in the text. The mass density for the initial conditions
is $\rho = 5$. The solid line denotes the 1-chip solution in Eq.~(\ref{soln11}).}
\label{Figure 2}
\end {figure}

\begin {figure}[htb!]
\centering
\includegraphics[width=12.0cm,height=10cm,angle=0]{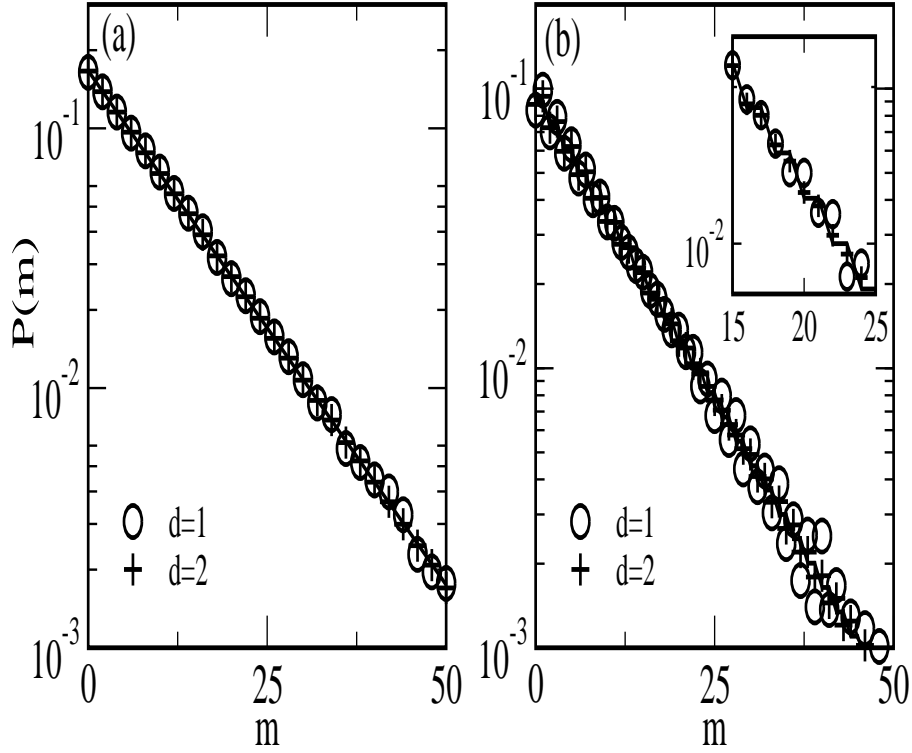}\\
\caption{Steady-state distributions for the 2-chip model, obtained from MC simulations. (a) Plot of $P(m)$ vs. $m$ on a linear-log scale, from MC simulations in
$d = 1,2$. The initial conditions were characterized by average density $\rho=10$, and 
$S_{\rm e}=1, S_{\rm o}=0$. The solid line denotes the solution in Eq.~(\ref{steady2})
with $s_{1}=s_{2}=\rho/(\rho + 2)$. (b) Analogous to (a), but the initial conditions for the
MC simulations have a mixture of both even and odd masses. The corresponding
parameter values are $\rho=9.5, S_{\rm e}=S_{\rm o}=1/2$. The solid line denotes the solution in
Eq.~(\ref{steady2}) with $s_{1}=(2\rho + 1)/(2\rho + 3)$, $s_{2}=(2\rho - 1)/(2\rho + 3)$.
The inset highlights the staircase structure of the probability distribution. }
\label{Figure 3}
\end {figure}

\begin {figure}[htb!]
\centering
\includegraphics[width=12.0cm,height=10cm,angle=0]{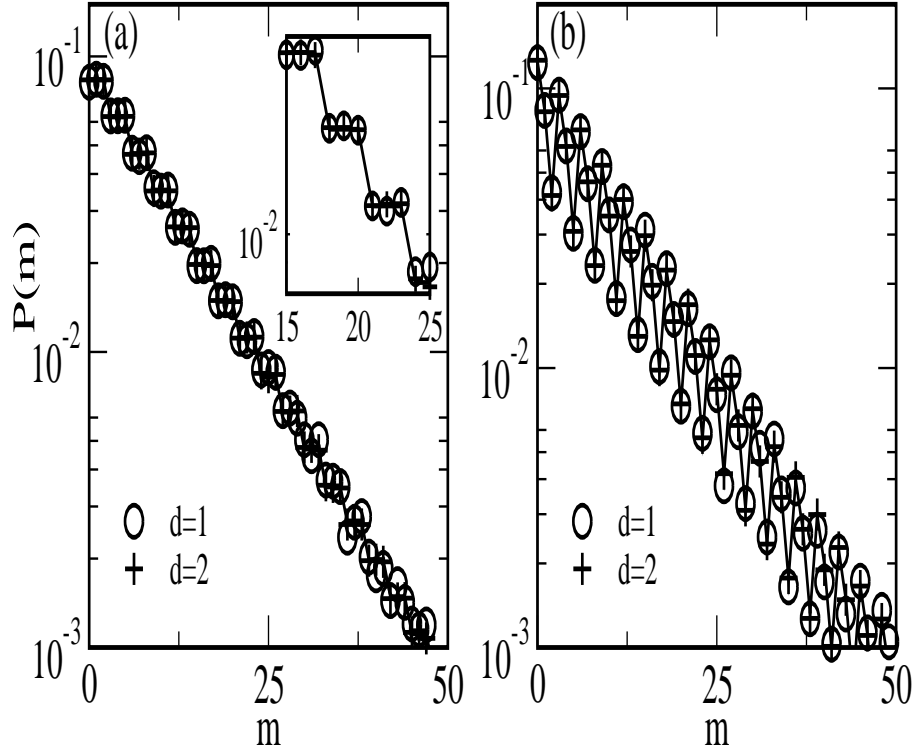}\\
\caption{Analogous to Fig.~3, but for the 3-chip model. (a) Plot of
$P(m)$ vs. $m$ from MC simulations with $P(9,0)=P(10,0)=P(11,0)=1/3$, so that $\rho=10$.
The solid line denotes the result in Eq.~(\ref{steadym}) with $s_{1}$, $s_{2}$ and
$s_{3}$ calculated from Eq.~(\ref{3chs}) as $s_1=0.917,s_2=0.833,s_3=0.75$. (b) Plot of
$P(m)$ vs. $m$ for initial conditions with $P(9,0)=1/2, P(10,0)=1/3, P(11,0)=1/6$, so
that $\rho \simeq 9.67$. The solid line denotes the result in
Eq.~(\ref{steadym}) with $s_{1}=0.875,s_{2}=0.792,s_{3}=0.75$.}
\label{Figure 4}
\end {figure}

\end{document}